\documentclass[referee,useAMS,usenatbib,usegraphicx]{mn2e}
\usepackage{amssymb}

\newcommand{\cG}{{\cal G}}
\newcommand{\cI}{{\cal I}}
\newcommand{\cH}{{\cal H}}
\newcommand{\cL}{{\cal L}}
\newcommand{\intp}{\int_{p_1}^{p_2}}
\newcommand{\intx}{\int_{x_1}^{x_2}}

\title[Shock acceleration of electrons in the presence of synchrotron losses] 
{Shock acceleration of electrons in the presence of synchrotron losses: I. test particle theory}
\author[Pasquale Blasi]{Pasquale Blasi$^{1}$\thanks{E-mail: blasi@arcetri.astro.it}\\ 
$^{1}$INAF-Osservatorio Astrofisico di Arcetri, Largo E. Fermi, 5, 50125, Firenze, Italy}

\begin{document}

\date{Accepted ----. Received -----}


\maketitle

\label{firstpage}

\begin{abstract}
We discuss a semi-analytical solution of the transport equation for electrons at a non-relativistic shock in the presence of synchrotron energy losses. We calculate the spectrum of accelerated (test) particles at any point upstream and downstream of the shock for an arbitrary diffusion coefficient  and we specialize the results to three cases: 1) diffusion constant in momentum ($D(p)=D_{0}$), 2) Bohm diffusion ($D(p)\propto p$), and 3) Kolmogorov diffusion ($D(p)\propto p^{1/3}$). 
Of special importance is the determination of the shape of the cutoff in the electron spectrum which depends on the diffusion properties felt by particles in the shock region. The formalism can be generalized to the case of a shock with an upstream precursor induced by the dynamical reaction of accelerated particles. 
\end{abstract}


\section{Introduction}

Particle acceleration of charged particles in the presence of losses has been subject of much investigation in the past, but in recent times its importance has been revived by the detection of non-thermal radiation from supernova remnants (SNRs) that, directly or indirectly, can be attributed to accelerated electrons in strong magnetic fields or intense radiation fields. Two instances can help stressing this point: observations carried out with the Chandra, XMM and Suzaku telescopes have shown that the non-thermal X-rays from SNRs are the result of synchrotron emission of highly relativistic electrons in magnetic fields which most likely are amplified with respect to the interstellar fields by a factor $\sim 100$ \cite[]{bamba03,bamba05,laz03,laz04,vink03,ballet}. In some cases the spectra of this radiation show a pronounced cutoff corresponding to the maximum energy of the radiating electrons. The spectrum of the X-rays and especially the shape of the cutoff carries precious information on the acceleration process and on the possible shock modification induced by the dynamical reaction of accelerated particles. Moreover, the morphology of the X-ray emission may lead to a direct estimate of the magnetic field strength in the shock region, which in turn affects particle diffusion. 

The magnetic field strength and topology determines the possibility for hadronic cosmic rays to reach the knee energy \cite[]{caprioli2007}. In turn the magnetic field can be generated by high energy cosmic rays if they excite streaming instabilities \cite[]{bell78,bell2004}. Electrons are affected by the magnetic field through diffusion and energy losses: both these effects may produce visible features in the cutoff region of the electron spectrum and eventually manifest themselves in the spectrum of the emitted radiation. 

Here we discuss the process of test-particle acceleration at a non-relativistic shock in the presence of synchrotron losses for three choices of the diffusion coefficient: 1) diffusion constant in momentum ($D(p)=D_{0}$), 2) Bohm diffusion ($D(p)\propto p$), and 3) Kolmogorov diffusion ($D(p)\propto p^{1/3}$). The transport equation at the shock surface is transformed in a simple equation to be solved iteratively and returns the spectrum of accelerated particles at the shock, while the spectrum at any location upstream and downstream is determined analytically through the proper Green functions.

Clearly this is not the first calculation of the process of electron acceleration at shocks, but previous methods that one can find in the literature are either numerical (solution of the transport equation through finite differences schemes) or analytical but applied to specific choices of the diffusion coefficient ($D(p)$ constant in momentum) or to limited regions of momentum space. For instance \cite{webb} used a Green function method to solve analytically the case of constant diffusion coefficient. The same case was then investigated by \cite{webbfritz} in the context of the so-called generations method, an elegant generalization of the stochastic method of \cite{bell78} for accelerated protons. \cite{zira} found an analytical solution for the case of Bohm diffusion but limited to particle momentum larger than the maximum momentum. The authors proposed a fitting formula, derived however from a numerical (finite differences) calculation, in order to interpolate the low energy and high energy regimes. The work by \cite{zira} is particularly important since it emphasizes the fact that in the case of Bohm diffusion, the spectrum of electrons has a super-exponential cutoff. This has obvious consequences in the comparison with observations of X-rays due to synchrotron emission of high energy electrons in the cutoff region. A numerical investigation of the problem, also including inverse Compton losses in the Klein-Nishina regime, was presented by \cite{vannoni}. In addition to these approaches that aimed at solving the exact transport equation, there are also attempts to face the problem of electron acceleration within leaky-box models \cite[]{boxdrury,boxstanev}. A generic prediction of these models is the appearance of pronounced pile-ups in the spectra of accelerated particles when the compression factor is larger than 4. Such pronounced and sharp features are typically absent in more detailed calculations.

The method discussed here provides both the spectrum of accelerated particles and their spatial distribution upstream and downstream of the shock, which is important to determine the morphology of the synchrotron emission. The calculation can in principle be carried out for any diffusion coefficient as function of momentum.

In this paper we restrict ourselves to the test particle regime of electron acceleration at a non-relativistic shock. However, while doing so we will obtain some results that will later be used to
generalize the calculations to the case of cosmic ray modified shocks. 

The paper is structured as follows: in \S \ref{sec:green} we illustrate the general theoretical bases of the method and the procedure to solve the relevant equations. In \S \ref{sec:results} we discuss the implementation of the iterative procedure, and illustrate the results of the calculations in terms of spectrum of the accelerated particles and their spatial distribution around the shock. We conclude in \S \ref{sec:concl}.

\section{The transport equation and Green functions}
\label{sec:green}

The acceleration of electrons at a shock front is described by the following transport equation:
\begin{equation}
u\frac{\partial f}{\partial x} = \frac{\partial}{\partial
    x}\left[D(p)\frac{\partial 
    f(x,p)}{\partial x} \right] + \frac{1}{3} \frac{du}{dx} p
\frac{\partial f(x,p)}{\partial p} -
\frac{1}{p^2}\frac{\partial}{\partial p} \left[ \dot p p^2 f(x,p)
  \right] + Q(x,p),
\label{eq:transport}
\end{equation}
where $\dot p = -A p^2$ ($A=\frac{4}{3}\sigma_{T}U_{B}/(m_{e}c)^{2}$, with $U_{B}=B^{2}/8\pi$, $B$ the magnetic field, $\sigma_{T}$ the Thomson cross section, $m_{e}$ the electron mass) is the rate of momentum loss due to synchrotron emission, $f(x,p)$ is the distribution function of accelerated particles and $Q$ is the injection term, assumed to have the form:
\begin{equation}
Q(x,p)=\frac{Q_0}{4\pi p^2}\delta(p-p')\delta(x-x').
\end{equation}
In Eq. \ref{eq:transport} we assume that energy losses allow the system to reach stationarity ($\partial f/\partial t=0$). In the test-particle regime, there is no precursor, therefore as usual $du/dx=(u_2-u_1)\delta(x)$, and the $x$ axis is oriented from upstream ($-\infty\leq x\leq 0$) to downstream ($0\leq x\leq +\infty$).  

For our purposes it is useful to introduce the function $N(x,p)=4\pi p^2 f(x,p)$, so that the transport equation reads:
\begin{equation}
\frac{\partial}{\partial x}\left[u N - D(p)\frac{\partial N}{\partial
    x} \right] = \frac{\partial}{\partial p}\left[A p^2 N\right] + Q_0
\delta(p-p') \delta(x-x').
\label{eq:transportN}
\end{equation}

The solution can be found by requiring the continuity of the distribution function and by satisfying the boundary condition at the shock, which is easily found by integrating the full transport equation in a narrow region around the shock. We will return to this point later.

We now focus our attention on solving Eq. \ref{eq:transportN} in the downstream and upstream fluids separately. For this purpose, following \cite{webbfritz}, we multiply Eq. \ref{eq:transportN} by an arbitrary function $\cG(x,p;x',p')$ and integrate both terms for $x_1\leq x\leq x_2$ and $p_1\leq p\leq p_2$, where the extremes of integration will be defined later in a suitable way. This results in the following expression:$$
\intp dp \left[\cG (uN-D\frac{\partial N}{\partial x})\right]_{x_1}^{x_2}+
\intp dp \intx dx D\frac{\partial}{\partial x}\left[ N \frac{\partial
    \cG}{\partial x}\right] - \intx dx \left[ A p^2 N \cG
  \right]_{p_1}^{p_2} -
$$
\begin{equation}
-\intp dp \intx dx N \left\{ u\frac{\partial \cG}{\partial x} + D
\frac{\partial^2 \cG}{\partial x^2} - A p^2 \frac{\partial
  \cG}{\partial p} \right\} = 0,
\label{eq:toadjoint}
\end{equation} 
where the injection term is not present as long as we concentrate on the regions upstream and downstream of the shock and we assume that the injection occurs exactly at the shock surface.

Here we also made use of the identity
\begin{equation}
\frac{\partial N}{\partial x}\frac{\partial \cG}{\partial x} = 
\frac{\partial}{\partial x} \left[ N \frac{\partial \cG}{\partial
    x}\right] - N \frac{\partial^2 \cG}{\partial x^2}.
\end{equation}

We recall now that the function $\cG(x,p;x',p')$ was chosen to be an arbitrary function of its arguments. We have therefore the freedom to choose it in the way that is most useful for our purposes. We choose the function $\cG$ as the solution of the equation
\begin{equation}
u\frac{\partial \cG}{\partial x} + D
\frac{\partial^2 \cG}{\partial x^2} - A p^2 \frac{\partial
  \cG}{\partial p} = - \delta(x-x') \delta(p-p'),
\label{eq:adj}
\end{equation}
In the downstream fluid we choose $x_1=0$ and $x_2=+\infty$, while in the upstream fluid, $x_1=-\infty$ and $x_2=0$. In both cases $p_1=0$ and $p_2=\infty$. Eq. \ref{eq:adj} is sometimes called the {\it adjoint equation}. Using these conditions in Eq. \ref{eq:toadjoint} we obtain the two fundamental expressions:
\begin{equation}
N_1(x,p) = -\int_{p}^{\infty} dp' D_1(p') N_0(p') \left[\frac{\partial \cG_1(x',p';x,p)}{\partial x'}\right]_{x'\to 0} - 
\int_{p}^{\infty} dp' \cG_1(x',p';x,p)|_{x'\to 0} \phi_{1}(p')
\label{eq:N1}
\end{equation}
in the upstream region, and 
\begin{equation}
N_2(x,p) = \int_{p}^{\infty} dp' D_2(p') N_0(p') \left[\frac{\partial \cG_2(x',p';x,p)}{\partial x'}\right]_{x'\to 0} +\int_{p}^{\infty} dp' \cG_2(x',p';x,p)|_{x'\to 0} \phi_{2}(p') ,
\label{eq:N2}
\end{equation}
in the downstream region. In order to keep the notation as simple as possible we inverted all primed and unprimed variables, so that the physical quantities are all expressed as functions of unprimed variables. The indexes ``1'' and ``2'' refer respectively to quantities in the upstream and downstream fluid. $N_0(p)$ is the spectrum of accelerated particles at the shock location. Clearly the distribution function of the accelerated particles is continuous across the shock, namely $\lim_{x\to 0^{-}}N_{1}(x,p)=\lim_{x\to 0^{+}} N_{2}(x,p) = N_{0}(p)$.
The lower limit of integration in Eqs. \ref{eq:N1} and \ref{eq:N2} is $p$ instead of zero because of the definition of the function $\cG$, solution of Eq. \ref{eq:adj} (see Appendix). From the physics point of view, it is obvious that it must be so, because the contribution to the spectrum at a given momentum $p$ can only come from particles with larger momentum that lose energy through synchrotron losses. The functions $\phi_{i}(p)$ are defined as:
\begin{equation}
\phi_{1,2}(p) = u_{1,2}N_{0}(p) - D_{1,2}(p)\frac{\partial N(x,p)}{\partial x}|_{1,2},
\end{equation}
which only depend on quantities evaluated at the shock. 

It should be noted that Eqs. \ref{eq:N1} and \ref{eq:N2} define the spectrum at any location upstream and downstream in terms of $N_0(p)$ and $\phi$ once the functions $\cG_1$ and $\cG_2$ are known. The solution at the shock, $N_0(p)$, must be derived by using the boundary condition at the shock location. On the other hand, the Green function of the adjoint equation, $\cG$, is easily calculated (see Appendix) to be:
\begin{equation}
\cG(x',z';x,z) = \frac{z^2}{2\pi A}\sqrt{\frac{\pi}{\tau(z',z)}}
\exp\left[-\frac{\left( x'-x+(u/A)(z-z') \right)^2}{4\tau(z',z)} \right]. 
\label{eq:green}
\end{equation}
Here we introduced $z=1/p$ and the function $\tau$, which is defined as follows:
\begin{equation}
\tau(z',z) = \frac{1}{A} \int_{p}^{p'} dy \frac{D(y)}{y^2}.
\end{equation} 

The main difference between the approach described here and the approach of \cite{webb} is in the fact that our approach (previously mastered by \cite{webbfritz} for diffusion coefficient independent of momentum) allows one to write the distribution function $N(x,t)$ as an integral on momentum of the distribution function at the shock, $N_{0}(p)$, and the particle flux $\phi$ weighed with the appropriate Green functions, which are easy to calculate. This proves convenient from the point of view of adopting iterative techniques to solve the equations.

In Eq. \ref{eq:green} all dependence on the diffusion coefficient is contained in the function $\tau$, which can be calculated for any choice of $D(p)$. The Green function calculated here is the one which vanishes at upstream and downstream infinity, but not at the shock. One could have chosen the Green function that satisfies the boundary condition of vanishing at the shock surface and the equations for $N(x,p)$ would have been simpler. However the calculations of such Green function is simple and computationally convenient only for the case of constant diffusion coefficient (see for instance \cite[]{webbfritz}). 

The boundary condition at the shock is easily obtained by integration of the full transport equation around the shock surface, and reads:
\begin{equation}
D_2(p) \left[\frac{\partial N}{\partial x}\right]_2 -
D_1(p) \left[\frac{\partial N}{\partial x}\right]_1 +
\frac{1}{3} (u_2-u_1) p \frac{dN_0}{dp} - 
\frac{2}{3} (u_2-u_1) N_0 = 0.
\label{eq:boundary}
\end{equation}

We introduce now the two functions
$$
\cI_1(p) = \frac{D_{1}(p)}{u_{1}N_{0}(p)}\frac{\partial N}{\partial x}|_{x\to 0^{-}}=
$$
\begin{equation}
-\frac{D_1(p)}{u_1 N_0(p)} \lim_{x\to 0}
\int_p^\infty d \bar p  D_1(\bar p) N_0(\bar p) \cH_{1}
-\frac{D_1(p)}{N_0(p)} \lim_{x\to 0} 
\int_p^\infty d\bar p  N_0(\bar p) \cL_{1} \left[1-\cI_{1}(\bar p)\right]
\label{eq:I1}
\end{equation}
and
$$
\cI_2(p) = \frac{D_{2}(p)}{u_{2}N_{0}(p)}\frac{\partial N}{\partial x}|_{x\to 0^{+}}=
$$
\begin{equation}
\frac{D_2(p)}{u_2 N_0(p)} \lim_{x\to 0}
\int_p^\infty d\bar p  D_2(\bar p) N_0(\bar p) \cH_{2}
+\frac{D_2(p)}{N_0(p)} \lim_{x\to 0} 
\int_p^\infty d\bar p  N_0(\bar p) \cL_{2} \left[1-\cI_{2}(\bar p)\right]
\label{eq:I2}
\end{equation}
where 
\begin{equation}
\cH_{1,2} (x,z,z')= \frac{\partial}{\partial x}\left[\frac{\partial \cG}{\partial \bar x}\right]_{\bar x\to 0^{\mp}} = \frac{\cG_{1,2}(\bar x=0)}{2\tau} \left[ 1- \frac{(-x+(u_{1,2}/A_{1,2})(z-z'))^{2}}{2\tau} \right],
\end{equation}
\begin{equation}
\cL_{1,2} (x,z,z') = \frac{\partial \cG_{1,2}(\bar x=0)}{\partial x} = 
\frac{\cG_{1,2}(\bar x=0)}{2\tau}\left[ -x + \frac{u_{1,2}}{A_{1,2}} (z-z') \right].
\label{eq:boundary2}
\end{equation}

It is important to notice that in the limit in which losses are negligible ($\tau\to 0$) one has $\cI_1\to 1$ and $\cI_2\to 0$. This behavior is exactly what one would expect based on the standard test-particle solution of the transport equation in the absence of losses: a physically meaningful, stationary solution can in fact be obtained only if the distribution function is homogeneous downstream, namely $\frac{\partial N(x=0^+)}{\partial x}=0$. At the same time the boundary condition upstream of the shock reads $D_1\frac{\partial N(x=0^-)}{\partial x}=-u_1 N_0$. This is exactly what the equations above are confirming. On the other hand, when synchrotron losses become important and the limit $\tau\to 0$ does not apply, Eqs. \ref{eq:I1} and \ref{eq:I2} allow us to write the natural generalization of the jump conditions to the case with synchrotron losses. 

In terms of $\cI_{1,2}$, implicit functions of $N_0(p)$, one can rewrite the boundary condition equation in a way which is reminiscent of the standard boundary condition for the loss-less stationary test-particle case:
\begin{equation}
\frac{dN_0^k}{N_0^k} = \frac{dp}{p} 
\left\{ 
\frac{\frac{2}{3}(u_2-u_1) - u_2 \cI_2^{k-1}(p) +
  u_1\cI_1^{k-1}(p)}{\frac{1}{3}(u_2-u_1)}  
\right\}.
\label{eq:slope}
\end{equation}

In the loss-less case, $\cI_2\to 0$ and $\cI_1\to 1$, so that Eq. \ref{eq:slope} reduces to the equation for the slope of a power law, which would then be $s=(2u_2+u_1)/(u_2-u_1)=-(2+r)/(r-1)$ where $r=u_1/u_2$ is the compression factor at the shock. The apexes $k$ and $k-1$ which have been introduced in Eq. \ref{eq:slope} are motivated by the fact that we intend to solve this equation by iterations: at each iteration the functions $\cI_i$ are evaluated at the previous step. The method converges in a few iterations to the spectrum of particles at the shock location, $N_0(p)$. At that point, the distribution function at an arbitrary point upstream and downstream is fully determined through Eqs. \ref{eq:N1} and \ref{eq:N2}. The method is computationally much faster than any finite differences scheme of numerical integration of the transport equation.

\section{Maximum momentum}

In the presence of losses, the maximum momentum can be estimated by requiring equality of the acceleration time and the loss time, the latter weighed by the propagation times in the upstream and downstream regions. The condition of balance between losses and acceleration reads:
\begin{equation}
(\tau_{R,1}+\tau_{R,2})\left( \frac{dE}{dt}\right)_{acc} - \tau_{R,1}\left( \frac{dE}{dt}\right)_{loss,1}
- \tau_{R,2}\left( \frac{dE}{dt}\right)_{loss,2} = 0,
\end{equation}
where $\tau_{R}$ are the residence times in the upstream and downstream regions. If we use, following \cite[]{druryacc}, $\tau_{acc}=E/\left( \frac{dE}{dt}\right)_{acc}=\frac{3}{u_{1}-u_{2}}\left[ \frac{D_{1}(p)}{u_{1}}+ \frac{D_{2}(p)}{u_{2}} \right]$ as the acceleration time, one has the condition
\begin{equation}
\tau_{acc}(p)=\frac{\tau_{R,1}+\tau_{R,2}}{ \frac{\tau_{R,1}}{\tau_{loss,1}} + 
\frac{\tau_{R,2}}{\tau_{loss,2}}},
\end{equation}
where $\tau_{loss}=E/(dE/dt)_{loss}$ is the loss time in the upstream (1) and downstream (2) region. This condition defines the maximum momentum $p_{max}$ for any choice of the diffusion coefficient.

In the following we provide some useful expressions for the three choices of the diffusion coefficient used in the present calculation. Some more details are provided for the case of Bohm diffusion since it is often adopted for phenomenological applications, especially when the magnetic field is strongly turbulent. 

\subsection{The case of constant diffusion coefficient}

For a constant diffusion coefficient $D(p)=D_0$, we have

\begin{equation}
\tau(z,z')=\frac{D_0}{A} (z-z'). 
\end{equation}

In principle, the diffusion coefficient and the rate of losses can be different upstream and downstream. This generalization is trivial to implement if needed and reflects in using the appropriate constant values for $D_0$ and $A$. 

The maximum momentum is defined by the expression:
\begin{equation}
p_{max}=\frac{1}{3D_{0}Ar}\frac{u_{1}^{2}(r-1)}{r+1},
\end{equation}
where $r=u_{1}/u_{2}$ is the compression factor at the shock.

\subsection{The case of Bohm diffusion}
\label{sec:PmaxBohm}

In this case we consider explicitly the most general situation in which the diffusion coefficients and rate of losses are different upstream and downstream. The diffusion coefficient has the functional dependence $D_{1,2}(p)=K_{1,2} p$, where $K_{1,2}=(1/3) c^{2}/(q B_{1,2})$ with $c$ the speed of light, $q$ the charge of electrons and $B$ the magnetic field in the shock region. This is also the same field strength used to determine the rate of losses: $\dot p=-A_{1,2}p^{2}=-(4/3)\sigma_{T}U_{B}/(m_e c)^2 p^{2}=-(4/9) \frac{q^{4}}{m_{e}^{4}c^{6}}B_{1,2}^{2} p^{2}$. We also have that $B_{1}/B_{2}=\kappa<1$.

The function $\tau$ reads:
\begin{equation}
\tau_{1,2}(z,z')=\frac{K_{1,2}}{A_{1,2}} \ln \left( \frac{z}{z'}\right).  
\end{equation}

The general expression for the maximum momentum is then
\begin{equation}
p_{max}=\sqrt{\left(\frac{r-1}{3r}\right)\left(\frac{1+\kappa r^{2}}{1+\kappa r}\right)
\left(\frac{1}{1+\kappa^{-1}r^{2}}\right)}
\frac{u_{1}(m_{e} c)^{2}}{\sqrt{\frac{4}{27} q^{3} B_{1}}}.
\end{equation}

This value of the maximum momentum, though a reasonable estimate, is not to be interpreted as the exact momentum where a cutoff in the spectrum of accelerated particles should appear. From the physical point of view this can be expected since the residence times upstream and downstream, used to derive $p_{max}$, are defined as $D/u^{2}$ only when losses are unimportant, therefore by definition not at $p\sim p_{max}$. \cite{zira} found that for the case of Bohm diffusion the cutoff is located at 
\begin{equation}
p_{0}=\left(\frac{r-1}{3r}\right) \left( \frac{1}{1+\kappa^{1/2}}\right)
\frac{u_{1}(m_{e} c)^{2}}{\sqrt{\frac{2}{27} q^{3} B_{1}}},
\label{eq:p0}
\end{equation}
namely
\begin{equation}
\frac{p_{max}}{p_{0}} = 
\sqrt{\frac{3r}{2(r-1)}} \left( 1+\kappa^{1/2}\right)\left(\frac{1+\kappa r^{2}}{1+\kappa r}\right)^{1/2} \sqrt{\frac{1}{1+\kappa^{-1}r^{2}}}.
\label{eq:p0pmax}
\end{equation}
For $r=4$ and $\kappa=1$ one has $p_{max}/p_{0}=1.26$.

\subsection{The case of Kolmogorov diffusion coefficient}

In this case $D(p)=K_{0}p^{1/3}$ and 
\begin{equation}
\tau (z,z') = \frac{3K_{0}}{2A}\left( z^{2/3}-z'^{2/3}\right).
\end{equation}
The maximum momentum is easily found to be:
\begin{equation}
p_{max}=\left[\frac{1}{3K_{0}Ar}\frac{u_{1}^{2}(r-1)}{r+1}\right]^{3/4}.
\end{equation}

\section{Results}
\label{sec:results}

In this section we discuss the results of the calculations. We apply the method to the three choices of the diffusion coefficient discussed above. We investigate the spatial and momentum dependence of the spectra of accelerated particles and the shape of the cutoff region for the three cases. 

\subsection{Spectrum at the shock and shape of the cutoff}

\begin{figure}
\begin{center}
  \includegraphics[angle=0,scale=.45]{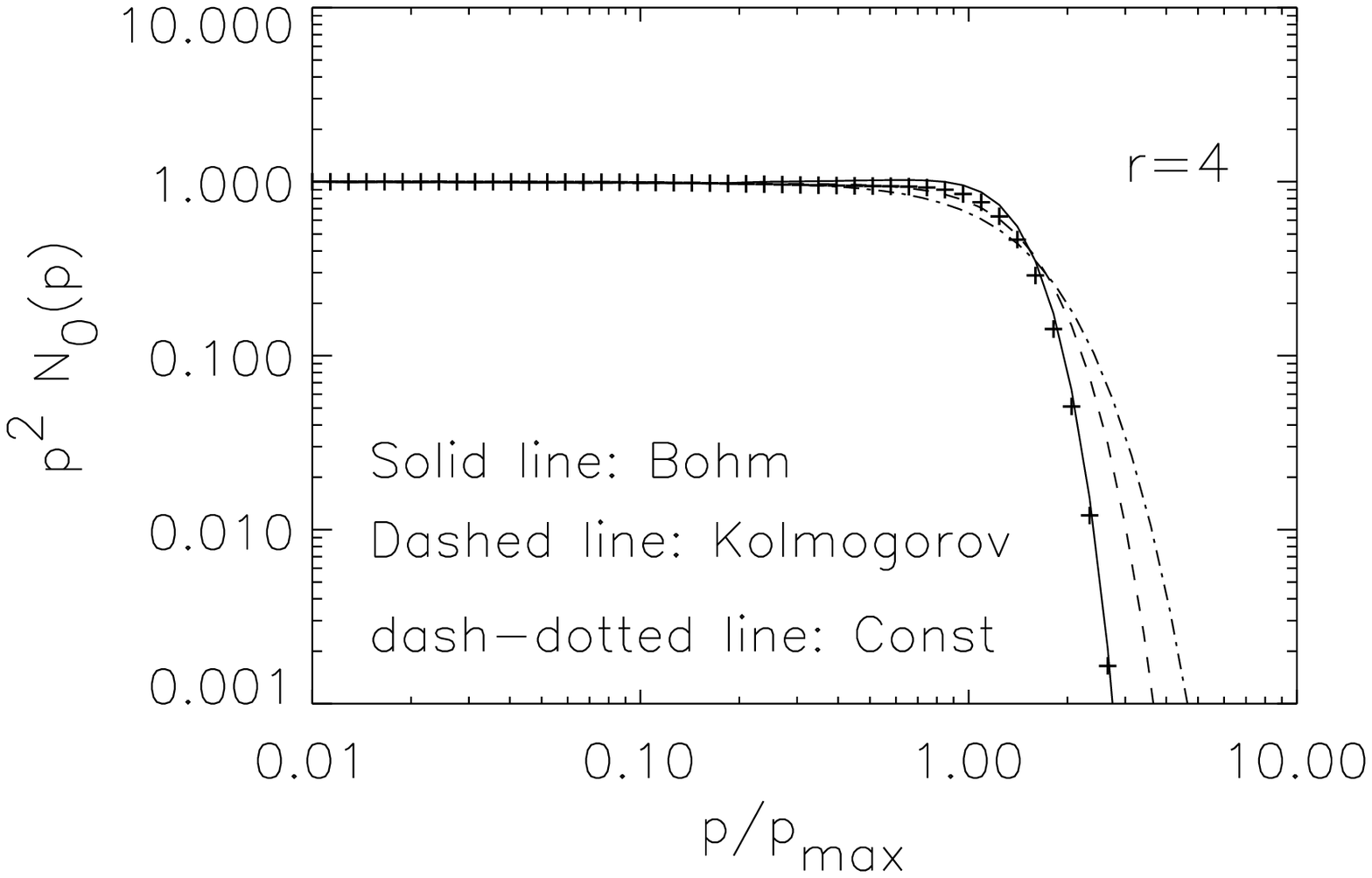}
  \includegraphics[angle=0,scale=.45]{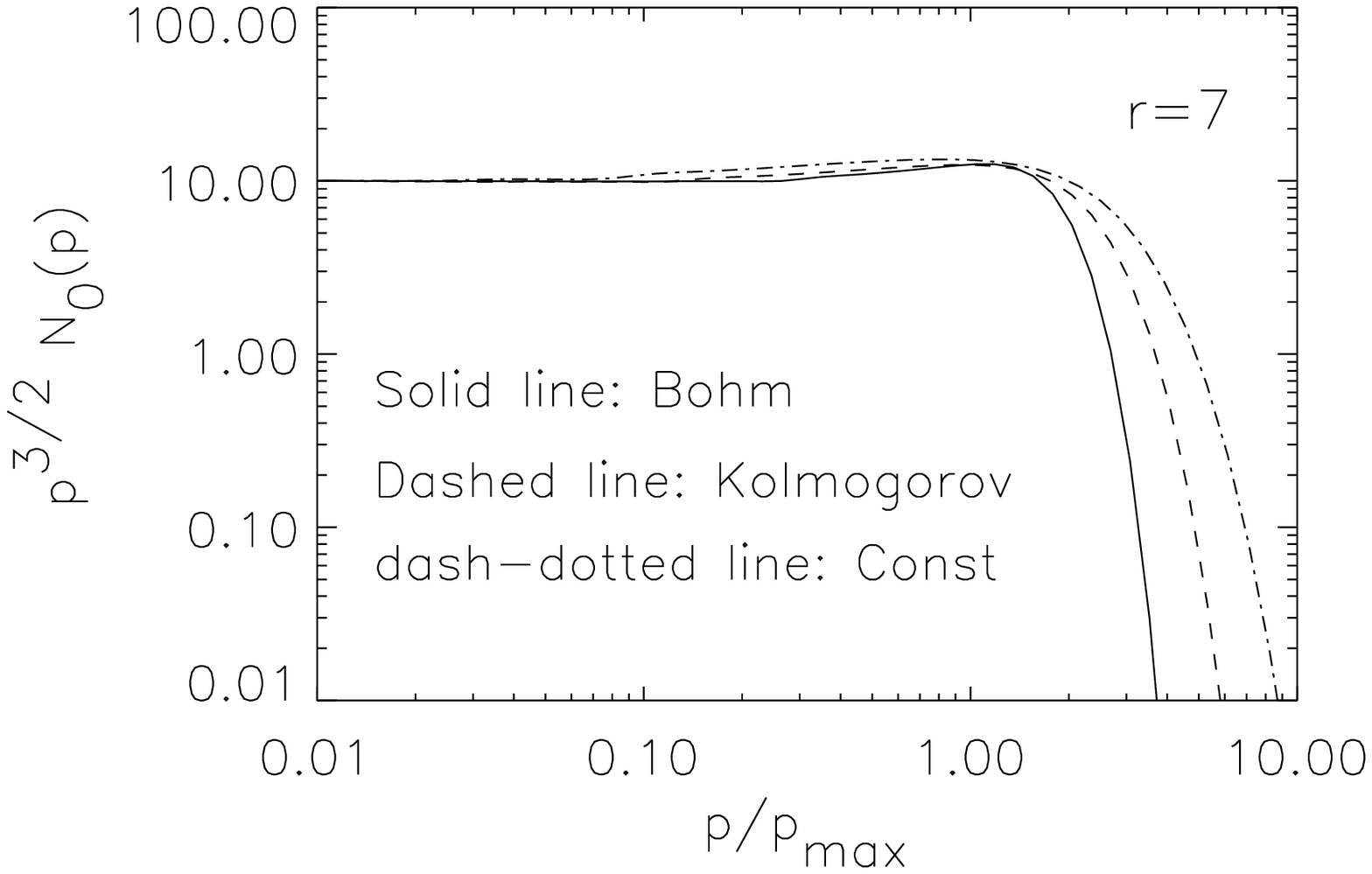}	
\caption{On the left (right) panel, spectrum of accelerated particles for $r=4$ ($r=7$) and for the three choices of the diffusion coefficient leading to the same value of $p_{max}$. The curves refer to constant diffusion coefficient (dash-dotted line), Kolmogorov diffusion (dashed line) and Bohm diffusion (solid line). The crosses in the left panel show the fit proposed by Zirakashvili \& Aharonian (2007).}
\label{fig:spectra}
\end{center}
\end{figure}

In the left panel of Fig. \ref{fig:spectra} we show the spectrum of accelerated particles at the shock location for $r=4$ and for the three choices of diffusion coefficient. The three diffusion coefficients are normalized in such a way to lead to the same maximum momentum $p_{max}$. The curves refer to $p^2 N_0(p)$ and constant diffusion coefficient (dash-dotted line), Kolmogorov diffusion (dashed line) and Bohm diffusion (solid line). 
The spectrum of accelerated particles has a shallower cutoff in the case with constant diffusion coefficient. The cutoff becomes sharper for diffusion coefficients with more pronounced dependence on particle momentum. For instance the shape of the cutoff for Bohm diffusion is $\propto \exp\left[-(p/p_{0})^{2}\right]$, where $p_{0}$ is defined in eq. \ref{eq:p0}. For the case of Bohm diffusion our resulting spectrum agrees well with the fitting formula proposed by \cite{zira}, based on an interpolation of an analytical approximation valid for $p>p_{max}$ and numerical calculations in the transition region (shown as crosses in Fig. \ref{fig:spectra}). It is of some importance that the position of the cutoff is not exactly at $p_{max}$ but rather at $p_{0}$ (see discussion in \S \ref{sec:PmaxBohm}). The discrepancy between the calculated spectra as presented here and as modeled by the fitting formula of \cite{zira} is typically fraction of percent, though at momenta $p\sim p_{max}$ it reaches  $\sim 10\%$, as visible in Fig. \ref{fig:spectra}. 

In the test particle limit the maximum compression factor which may be realized at the shock is $r=4$, leading to $N_0(p)\propto p^{-2}$. In cosmic ray modified shocks the effective total compression factor between upstream infinity and downstream can be much higher. For the sole purpose of illustrating the effect that may be expected in such cases, we consider here the case $r=7$. The spectra, again normalized by the spectrum that would be obtained in the absence of synchrotron
losses ($N_0(p)\propto p^{-3/2}$ in this case), are illustrated in the right panel of Fig. \ref{fig:spectra} for the three diffusion coefficients introduced above. At low momenta, $p\ll p_{max}$, where synchrotron losses are negligible, the standard test particle spectrum is reproduced. The most noticeable difference with the case $r=4$ is the well visible bump around
$p=p_{max}$ for all choices of diffusion coefficient. The bump is broader for constant diffusion coefficient and becomes gradually more prominent and narrower for stronger dependence of the diffusion coefficient on momentum.

\subsection{Spatial distribution of the accelerated particles}

\begin{figure}
\begin{center}
\includegraphics[angle=0,scale=.45]{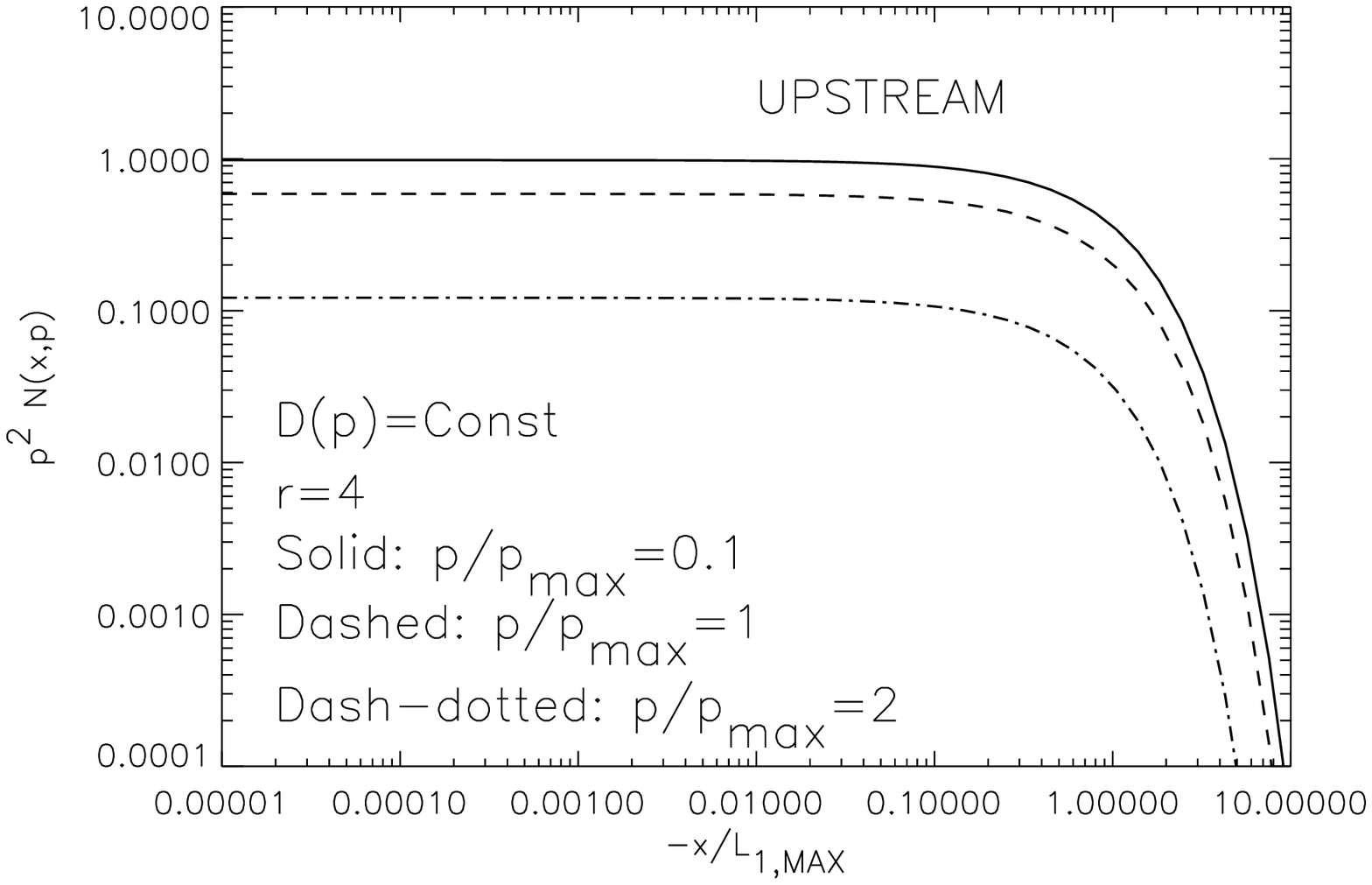}
\includegraphics[angle=0,scale=.45]{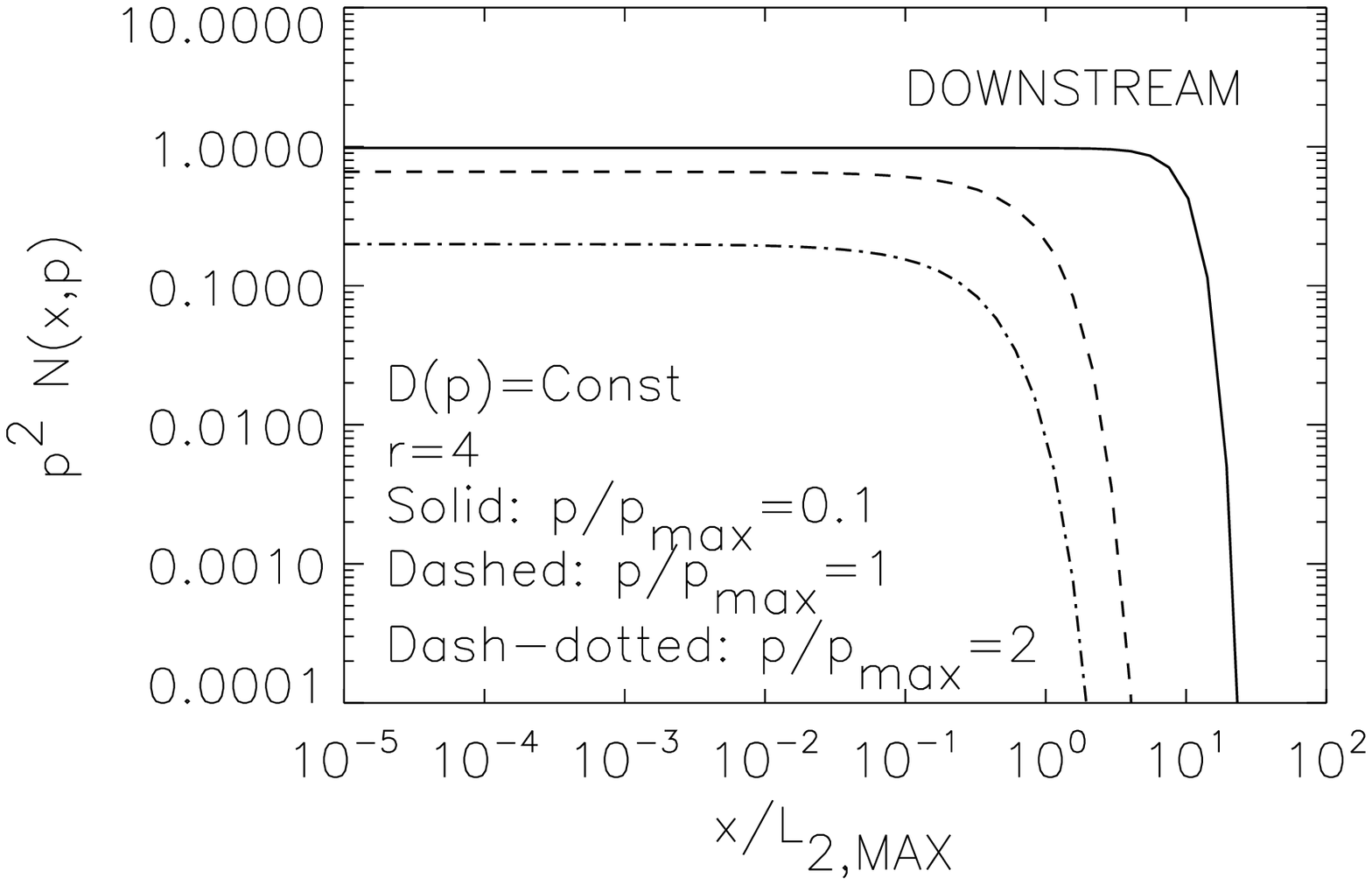}
\caption{Spatial distribution of accelerated electrons for constant diffusion coefficient and $r=4$. The left (right) panel refers to the upstream (downstream) region. The curves refer to $p=0.1 p_{max}$ (solid line), $p=p_{max}$ (dashed line) and $p=2 p_{max}$ (dash-dotted line).}
\label{fig:space4_const}
\end{center}
\end{figure}
The approach described above also allows one to determine the spatial distribution of the electrons in the downstream and upstream regions, through Eqs. \ref{eq:N1} and \ref{eq:N2}. This information is important for the determination of the spectrum of the radiation emitted by the accelerated electrons. 

In Fig. \ref{fig:space4_const} we show the spatial distribution of the electrons upstream (left panel) and downstream (right panel) for constant diffusion coefficient and for three values of the momentum of particles: $p=0.1 p_{max}$ (solid line), $p=p_{max}$ (dashed line) and $p=2 p_{max}$ (dash-dotted line). On the y-axis we plot $p^2 N(x,p)$. On the x-axis there is the distance from the shock in units of $L_{1,max}=D_{0}/u_{1}$ upstream and $L_{2,max}=u_{2}\tau_{loss,2}(p_{max})$ downstream. In all cases a compression factor $r=4$ has been adopted. 

It should be noticed that the diffusion length is independent of momentum in this case, therefore the size of the upstream region is basically independent of $p$ at low momenta (compare the dashed and solid lines). At momenta close to the maximum momentum, losses become important, but by definition of maximum momentum the acceleration time and the loss time are comparable at that point, therefore the size of the diffusion region in the upstream fluid is roughly the same also for $p\sim p_{max}$. At $p>p_{max}$ the density of particles is suppressed exponentially, as shown in Fig. \ref{fig:space4_const} (dash-dotted line).

In the downstream region the situation is more interesting: there are two relevant spatial scales, $x_{loss}^{adv}=u_2\tau_{loss}\sim 1/p$ and $x_{loss}^{diff}= \sqrt{2 D_0\tau_{loss}}\sim 1/p^{1/2}$. For a diffusion coefficient constant with momentum, the two spatial scales are equal at $p_*=u_2^2/(2AD_0)$. Comparing this with the maximum momentum, one obtains:
\begin{equation}
\frac{p_*}{p_{max}} = \frac{3}{2}\frac{r+1}{r(r-1)}.
\end{equation}
For $r=4$, $\frac{p_*}{p_{max}} = 5/8$, which implies that at $p<(5/8)p_{max}$ the size of the region where accelerated particles are located in the downstream fluid is of order $x_{loss}^{adv}(p)=u_2\tau_{loss}(p)$. This scaling with momentum can be qualitatively seen in the right panel of Fig. \ref{fig:space4_const}, by comparing the cases with $p=0.1 p_{max}$ and $p=p_{max}$. 

At small distances from the shock, $p^2 N(x,p)$ is spatially constant at low momenta, as it should be since synchrotron losses become negligible.  
\begin{figure}
\begin{center}
\includegraphics[angle=0,scale=.45]{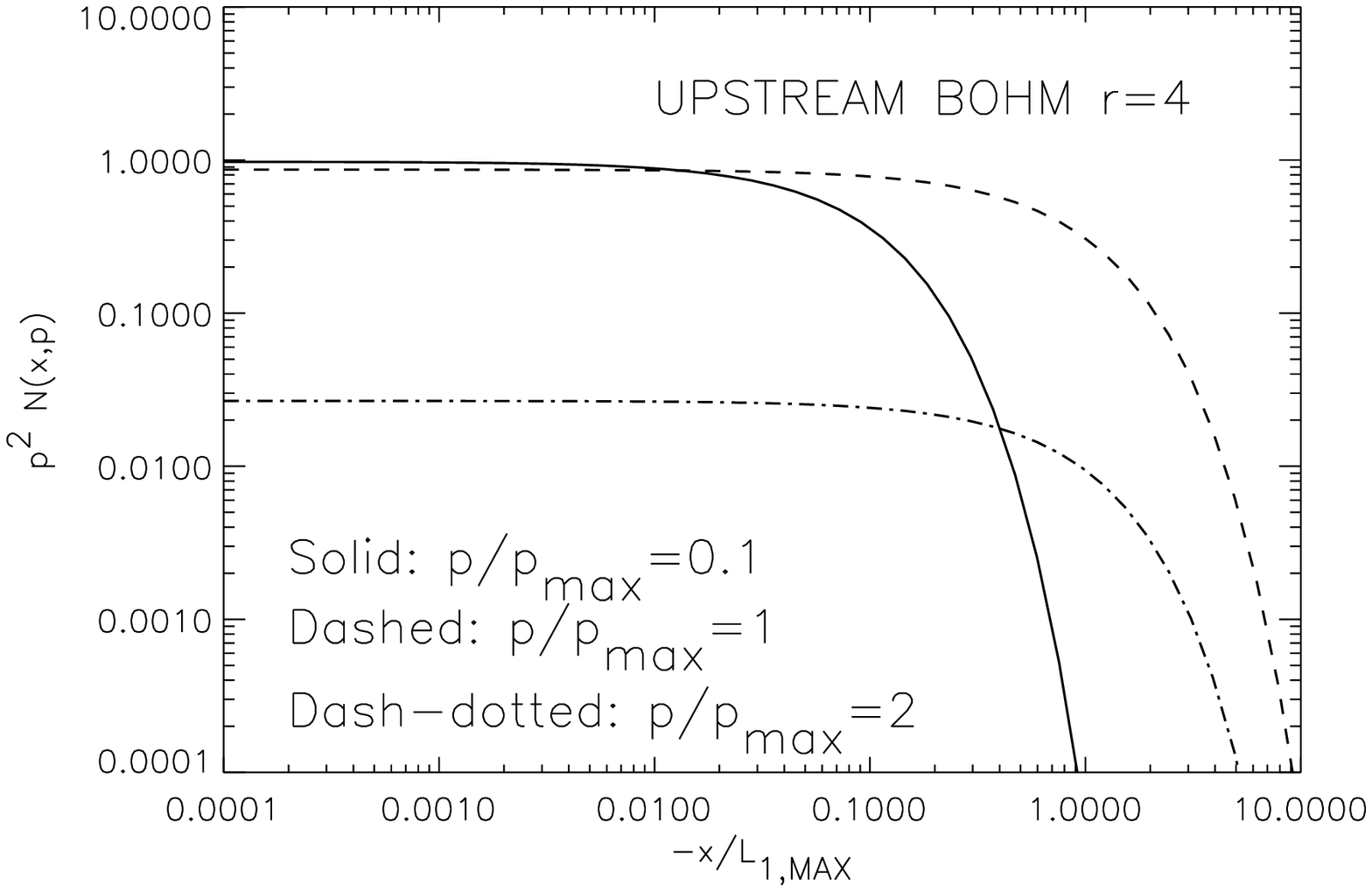}
\includegraphics[angle=0,scale=.45]{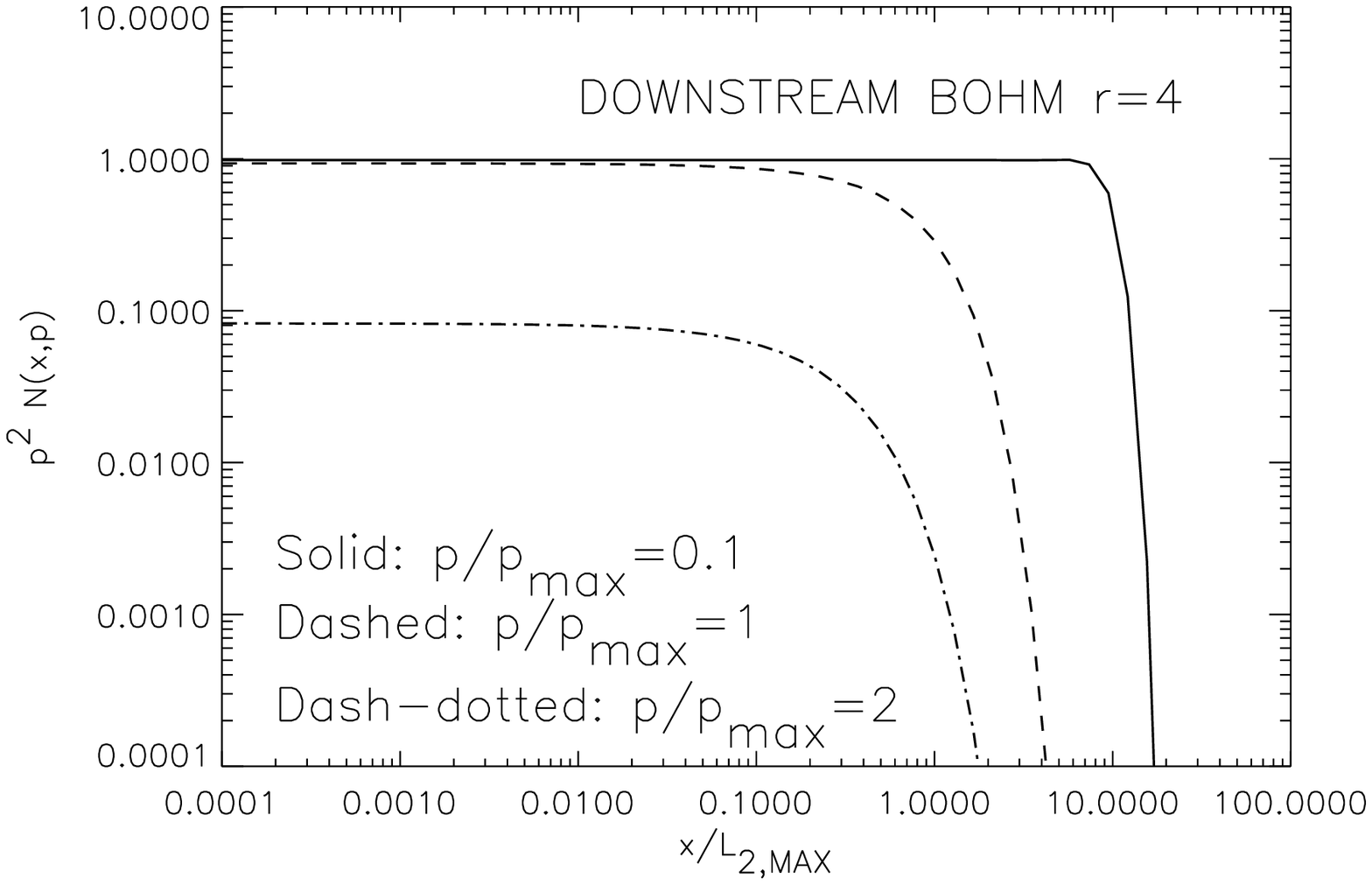}
\caption{Spatial distribution of accelerated electrons for a Bohm diffusion coefficient and $r=4$. The left (right) panel refers to the upstream (downstream) region. The curves refer to $p=2 p_{max}$ (dash-dotted line), $p=p_{max}$ (dashed line), $p=0.1 p_{max}$ (solid line).}
\label{fig:space4_bohm}
\end{center}
\end{figure}

\begin{figure}
\begin{center}
\includegraphics[angle=0,scale=.45]{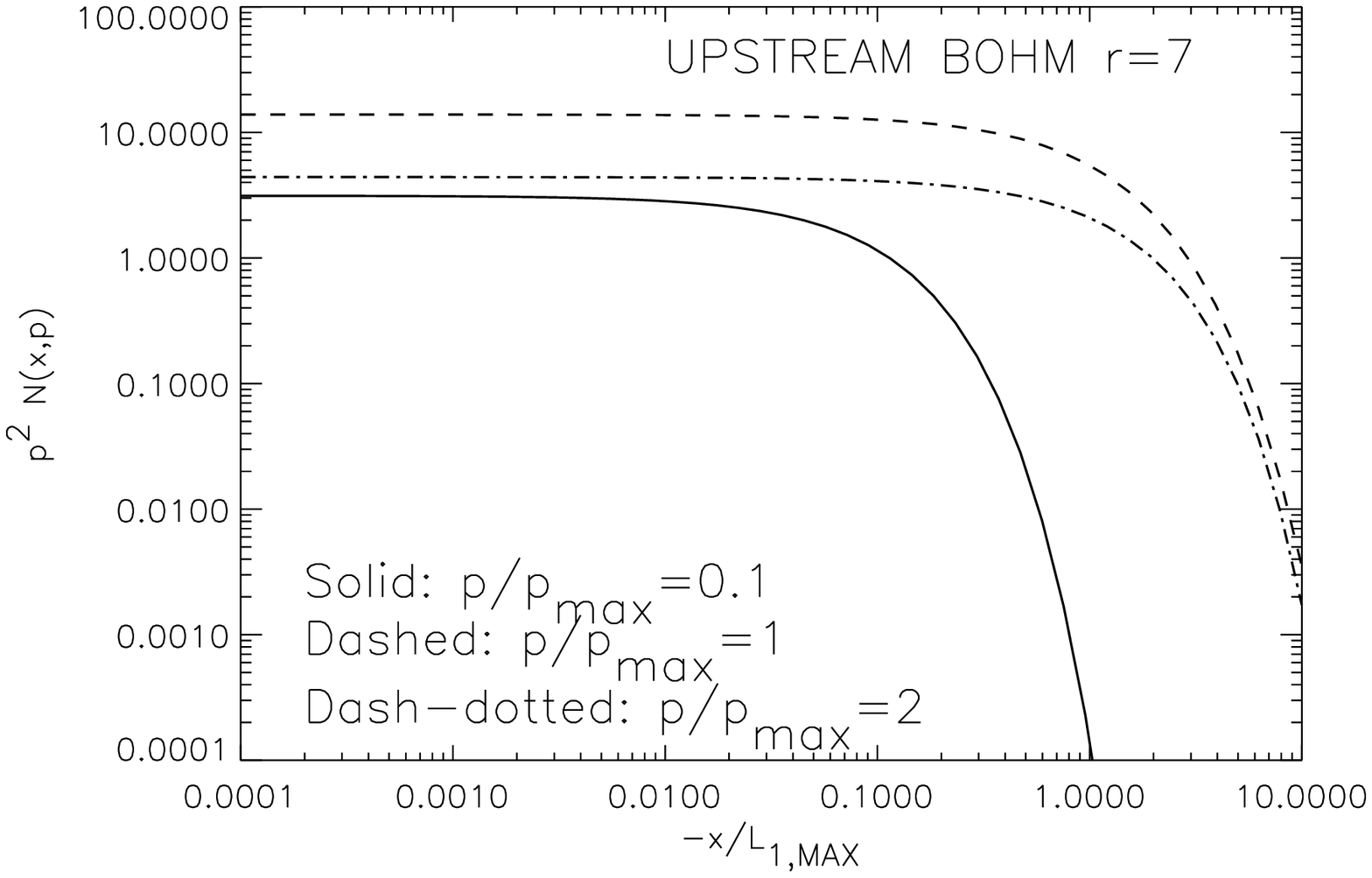}
\includegraphics[angle=0,scale=.45]{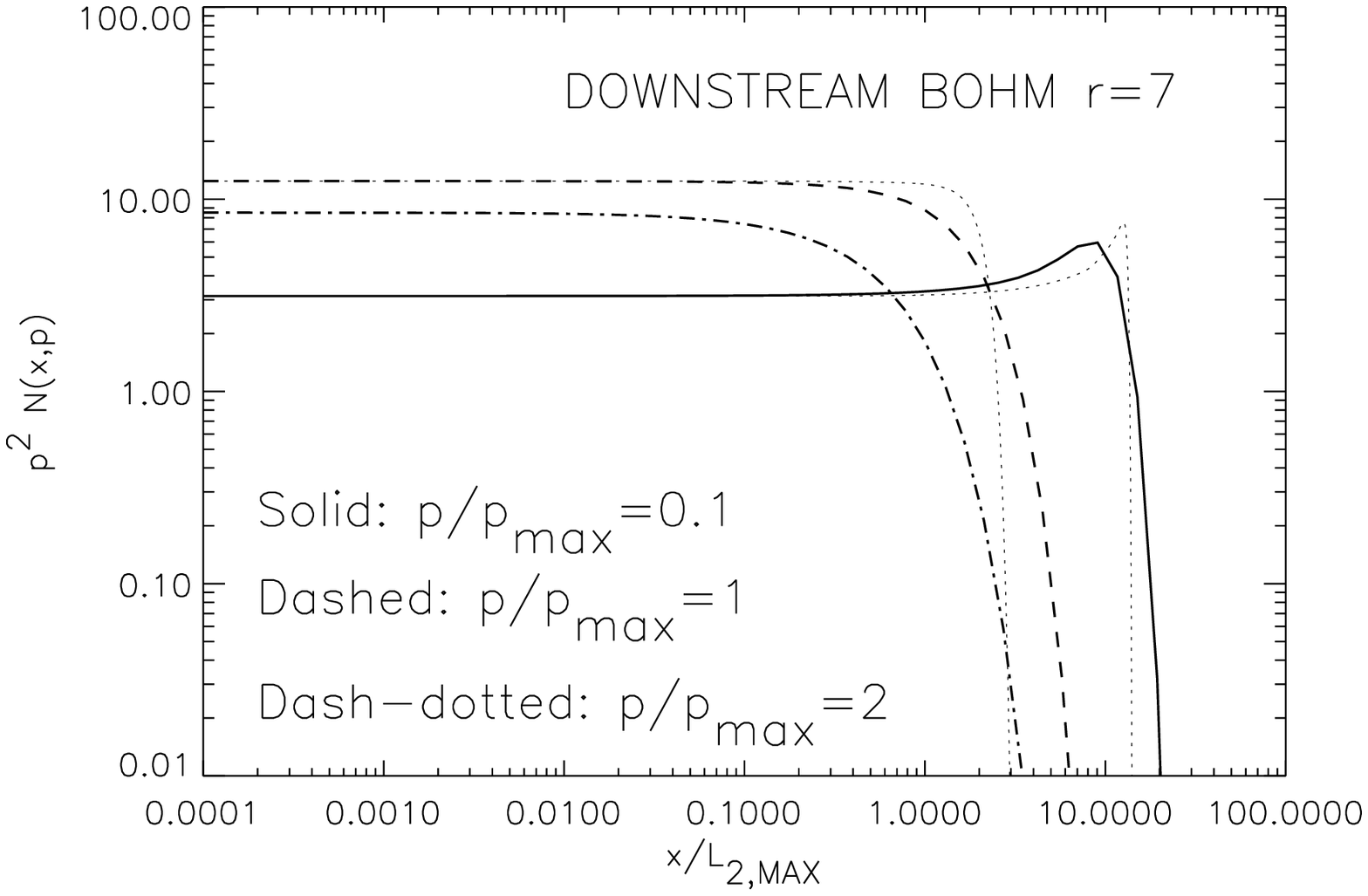}
\caption{Spatial distribution of accelerated electrons for a Bohm diffusion coefficient and $r=7$. The left (right) panel refers to the upstream (downstream) region. The curves refer to $p=2 p_{max}$ (dash-dotted line), $p=p_{max}$ (dashed line), $p=0.1 p_{max}$ (solid line). The thin dotted lines show the result of a qualitative argument for the development of bumps, as discussed in the text.}
\label{fig:space7_bohm}
\end{center}
\end{figure}

Our results on the spatial distribution for the case of Bohm diffusion are illustrated in Fig. \ref{fig:space4_bohm}. The distances upstream are normalized to $L_{1,max}=D(p_{max})/u_{1}$, while downstream $L_{2,max}=u_{2} \tau_{loss}(p_{max})$. The left (right) panel refers to the spatial distribution of accelerated particles ($p^2 N(x,p)$) in the upstream (downstream) region. The curves refer to different momenta of accelerated particles:  $p=0.1 p_{max}$ (solid line), $p=p_{max}$ (dashed line) and $p=2 p_{max}$ (dash-dotted line).

The main differences with the previous case are all due to the fact that now the diffusion coefficient depends on momentum. In the upstream region, the size of the diffusion region for $p<p_{max}$ scales linearly with the momentum, as clearly shown in the left panel of Fig. \ref{fig:space4_bohm}. At $p>p_{max}$ the synchrotron losses are faster than acceleration and the spatial size of the particle distribution starts decreasing.

The situation downstream is again more interesting. Introducing once again the two distances $x_{loss}^{adv}=u_2\tau_{loss}(p)\sim 1/p$ and $x_{loss}^{diff}= \sqrt{2 D_{0,B} p\tau_{loss}(p)}$ (constant in momentum), one sees that the spatial size of the downstream region for particles of momentum $p$ is given by  $x_{loss}^{adv}=u_2\tau_{loss}(p)\propto 1/p$ for all momenta such that $p/p_{max}<\sqrt{3(r+1)/(2r(r-1))}$. For $r=4$, the condition becomes $p/p_{max}< 0.8$. The scaling with $1/p$ can again be identified in the right panel of Fig. \ref{fig:space4_bohm}. Moving to lower and lower energies the drop in the spatial density of accelerated particles at $x>L_{2,max}$ becomes increasingly sharper.

In Fig. \ref{fig:space7_bohm} we plot the results for compression factor $r=7$. The main difference with the previous cases is the appearance of pile-ups in the spatial distribution of low energy accelerated particles in the downstream plasma. This feature is due to the energy loss of particles at $p\sim p_{max}$ around the bump which appears right before the cutoff (see Fig. \ref{fig:spectra}).
The bumps become more pronounced at lower energies while being absent at $p\gtrsim p_{max}$.

The bumps are not artifacts of the calculation, as can be shown rather easily at least qualitatively: if advection is the only relevant process, then $x=u_{2}t$ downstream, and one can write $dp/dx=-(A/u_{2}) p^{2}$. Conservation of the number of particles between the shock (where the particle density is $N_{0}(p)$ and the location $x$, where the particle density is $N(p,x)$, then easily leads to
\begin{equation}
N(p,x)dp=N_{0}(\bar p)d \bar p \to N(p,x) = \frac{N_{0}(\bar p)}{\left( 1-\frac{Axp}{u_{2}} \right)^{2}} \propto \frac{p^{-\gamma}}{\left( 1-\frac{Axp}{u_{2}} \right)^{2-\gamma}},
\label{eq:Npx}
\end{equation}
where $\bar p=p/(1-Axp/u_{2})$ is the momentum that a particle is produced with at the shock in order to reach the location $x$ downstream with momentum $p$, as calculated from the rate of momentum losses. The last step in Eq. \ref{eq:Npx} holds for the case in which $N_{0}(p)\propto p^{-\gamma}$. In this case one can easily show that a spike develops at values of $x$ such that the denominator vanishes, provided $\gamma<2$ (namely $r>4$). In reality these spikes are smoothened by the presence of a cutoff in $N_{0}(p)$ and even more important by the fact that at sufficiently high momenta diffusion prevails upon advection. In a qualitative way one could describe this effect by writing $x=u_{2}t + \sqrt{2D(p)t}$ and repeat the argument above. The result is not completely analytical but it requires the numerical integration of a simple differential equation which results in the dotted thin lines shown in the right panel of Fig. \ref{fig:space7_bohm}. These qualitative results show how the bumps in fact are expected for $r>4$ and for low enough momenta, while they disappear for $p\gtrsim p_{max}$. 

\subsection{Spectra at different locations upstream and downstream}

\begin{figure}
\begin{center}
\includegraphics[angle=0,scale=.45]{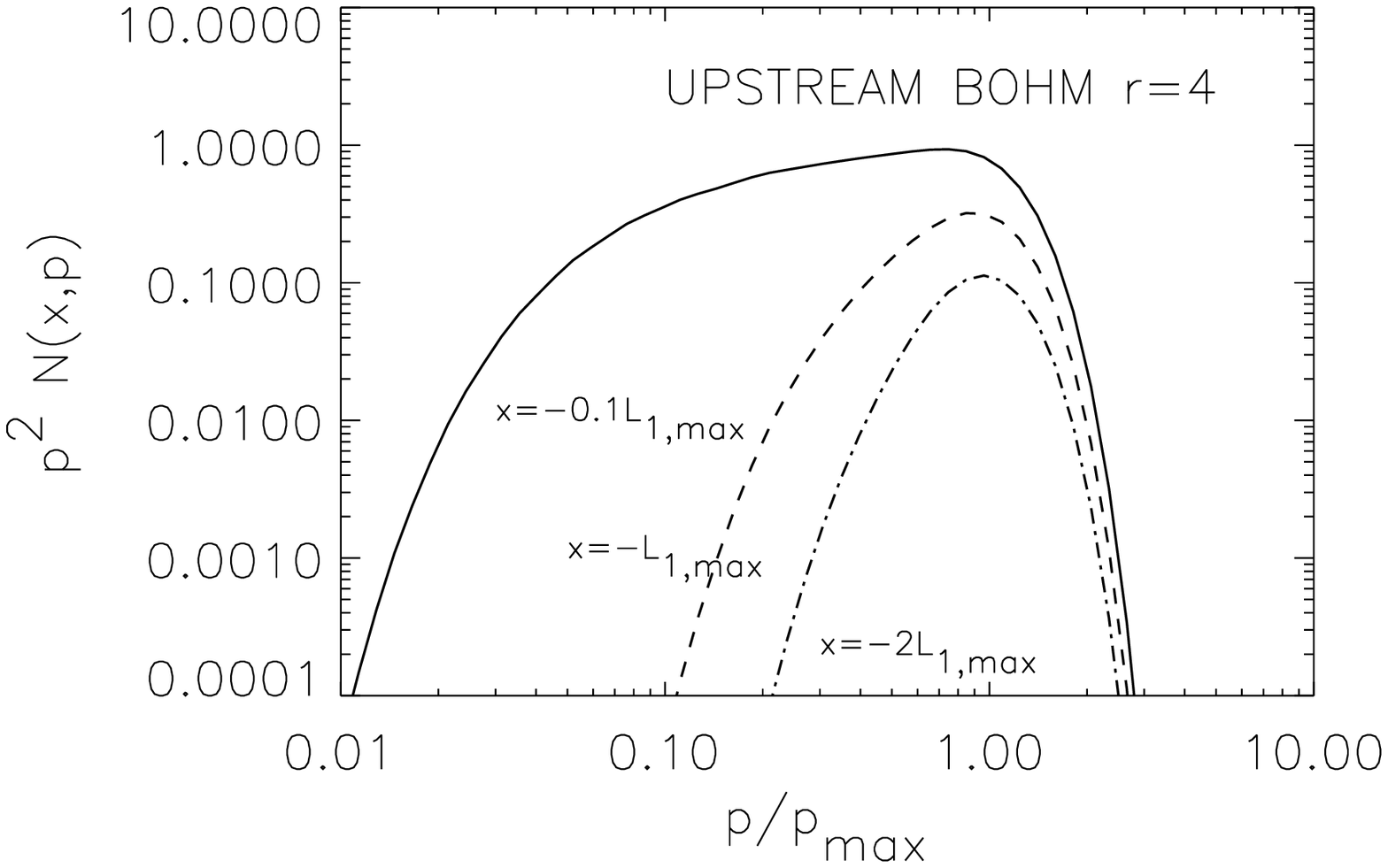}
\includegraphics[angle=0,scale=.45]{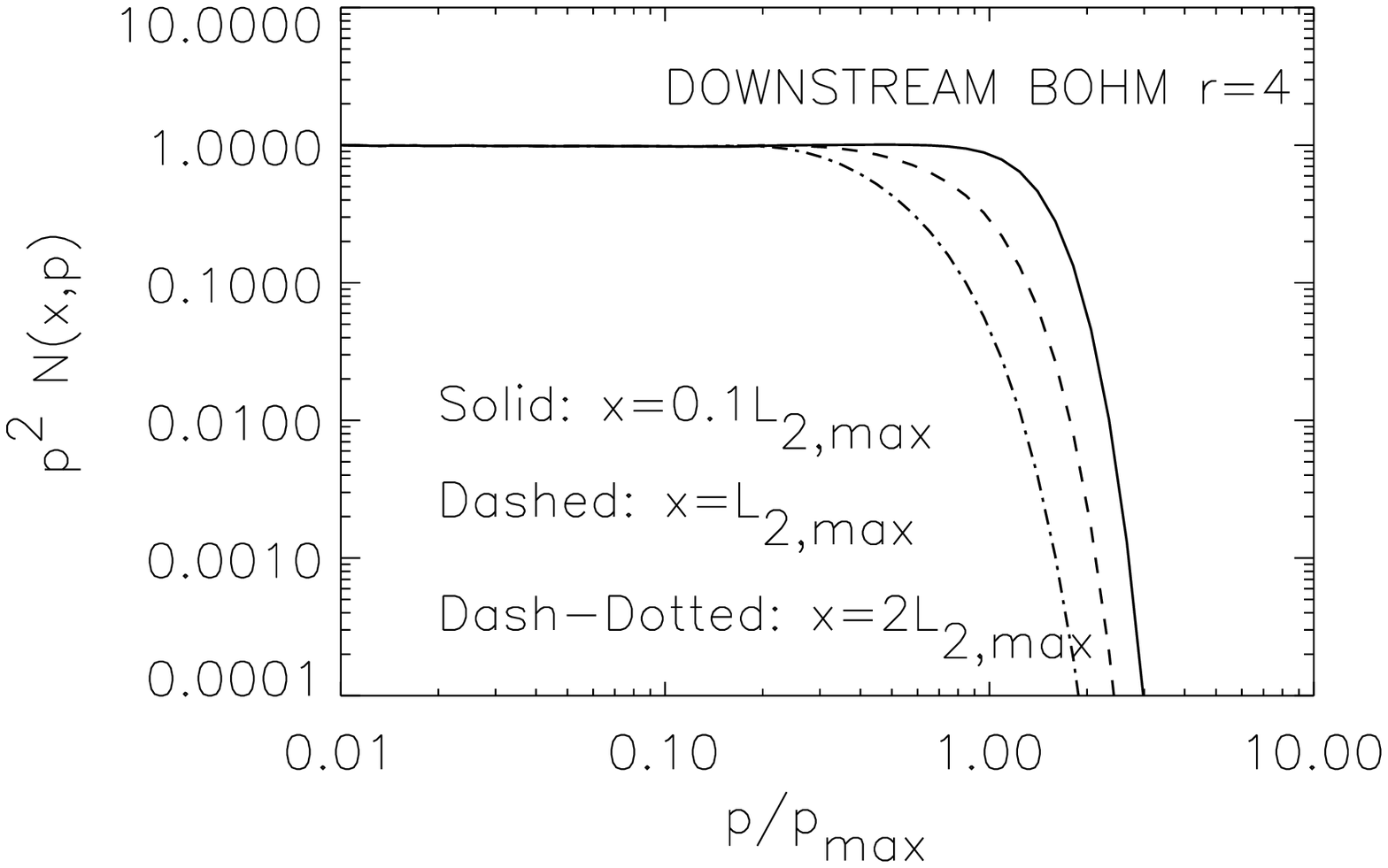}
\caption{Spectra of accelerated electrons upstream (left panel) and downstream (right panel) for a Bohm diffusion coefficient and $r=4$. The curves are for $x=-0.1L_{1,max}$ (solid line), $x=-L_{1,max}$ (dashed line) and $x=-2 L_{1,max}$ upstream and $x=0.1L_{2,max}$ (solid line), $x=L_{2,max}$ (dashed line) and $x=2 L_{2,max}$ downstream.}
\label{fig:spectradistance}
\end{center}
\end{figure}

In Fig. \ref{fig:spectradistance} we plot the spectrum of accelerated particles at different distances upstream (left panel) and downstream (right panel) for Bohm diffusion. In the left panel the spectrum is plotted at $x=-0.1L_{1,max}$ (solid line), $x=-L_{1,max}$ (dashed line) and $x=-2 L_{1,max}$ (dash-dotted line), where $L_{1,max}=D(p_{max})/u_{1}$. In the upstream plasma we can see the expected trend of lower energy particles to be confined closer to the shock surface. At increasingly larger distances from the shock only high energy particles are present. On the other hand at $p>p_{max}$ energy losses become important and the absolute number of particles drops down. 

In the right panel the spectrum is plotted at $x=0.1L_{2,max}$ (solid line), $x=L_{2,max}$ (dashed line) and $x=2 L_{2,max}$ (dash-dotted line) with $L_{2,max}=u_{2}\tau_{loss}(p_{max})$.
In the downstream plasma the main effect to be noticed is that the spectrum is truncated at lower energies while moving further downstream, as a result of synchrotron energy losses that cut out high energy particles from the spectrum. 

\subsection{Spatially integrated spectra of accelerated particles}

Although X-ray imaging has allowed us to observe the morphology of the radiation emitted by accelerated particles, it remains of interest to measure the spatially integrated emission from the shock region, which often is the only quantity we can actually measure. In this section we discuss our results for the spatially integrated spectra of accelerated particles, which have a behavior similar to that of the integrated emission. The spatially integrated spectra can be obtained by integrating Eqs. \ref{eq:N1} and \ref{eq:N2} in $x$ in the appropriate regions upstream and downstream, namely:
\begin{equation}
S_{1}(p) = \int_{-\infty}^{0} dx N_{1}(x,p)~~~~~~~~~~~S_{2}(p) = \int_{0}^{\infty} dx N_{2}(x,p).
\end{equation}
After calculating analytically the integral over $x$ we obtain:
$$
S_{1}(p) = \frac{z^{2}}{2A_{1}\pi^{1/2}}\int_{p}^{\infty} dp' D_{1}(p') N_{0}(p')\times 
\frac{1}{\tau^{1/2}} \exp\left[ - \frac{u_{1}^{2} (z'-z)^{2}}{4 A_{1}^{2} \tau}\right] -
$$
\begin{equation}
-\frac{z^{2}u_{1}}{2A_{1}}\int_{p}^{\infty} dp' N_{0}(p') \left\{ 1+ Erf\left[ \frac{u_{1}(z'-z)}{2A_{1}\tau^{1/2}} \right]\right\} \phi_{1}(p').
\end{equation}

Proceeding in the same way, the integral over the downstream region reads:
$$
S_{2}(p) = \frac{z^{2}}{2A_{2}\pi^{1/2}}\int_{p}^{\infty} dp' D_{2}(p') N_{0}(p')\times 
\frac{1}{\tau^{1/2}} \exp\left[ - \frac{u_{2}^{2} (z'-z)^{2}}{4 A_{2}^{2} \tau}\right] -
$$
\begin{equation}
-\frac{z^{2}u_{2}}{2A_{2}}\int_{p}^{\infty} dp' N_{0}(p') \left\{ 1-Erf\left[ \frac{u_{2}(z'-z)}{2A_{2}\tau^{1/2}} \right]\right\} \phi_{2}(p').
\end{equation}

Here $Erf(y)$ is the error function.

\begin{figure}
\begin{center}
  \includegraphics[angle=0,scale=.45]{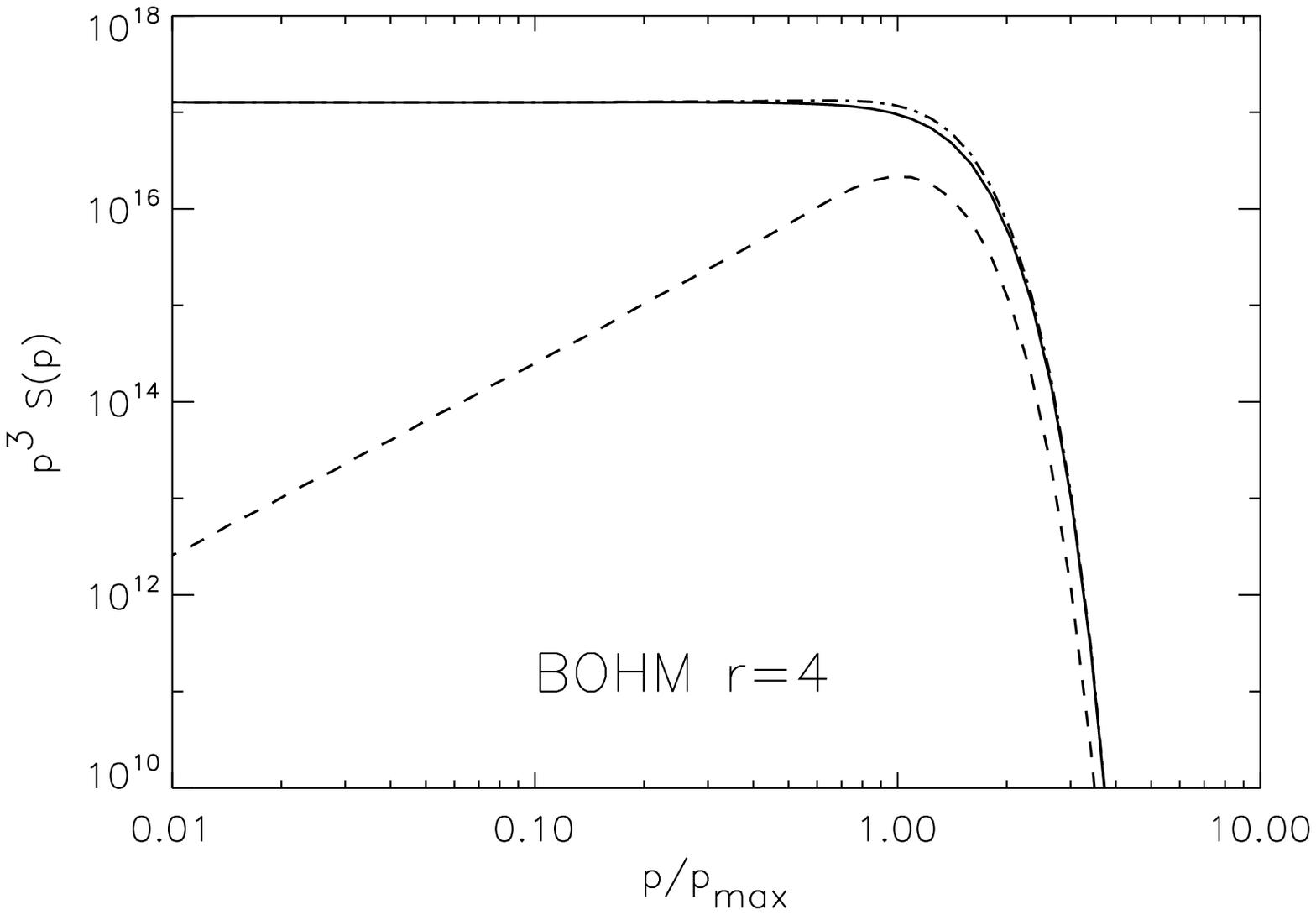}
   \includegraphics[angle=0,scale=.45]{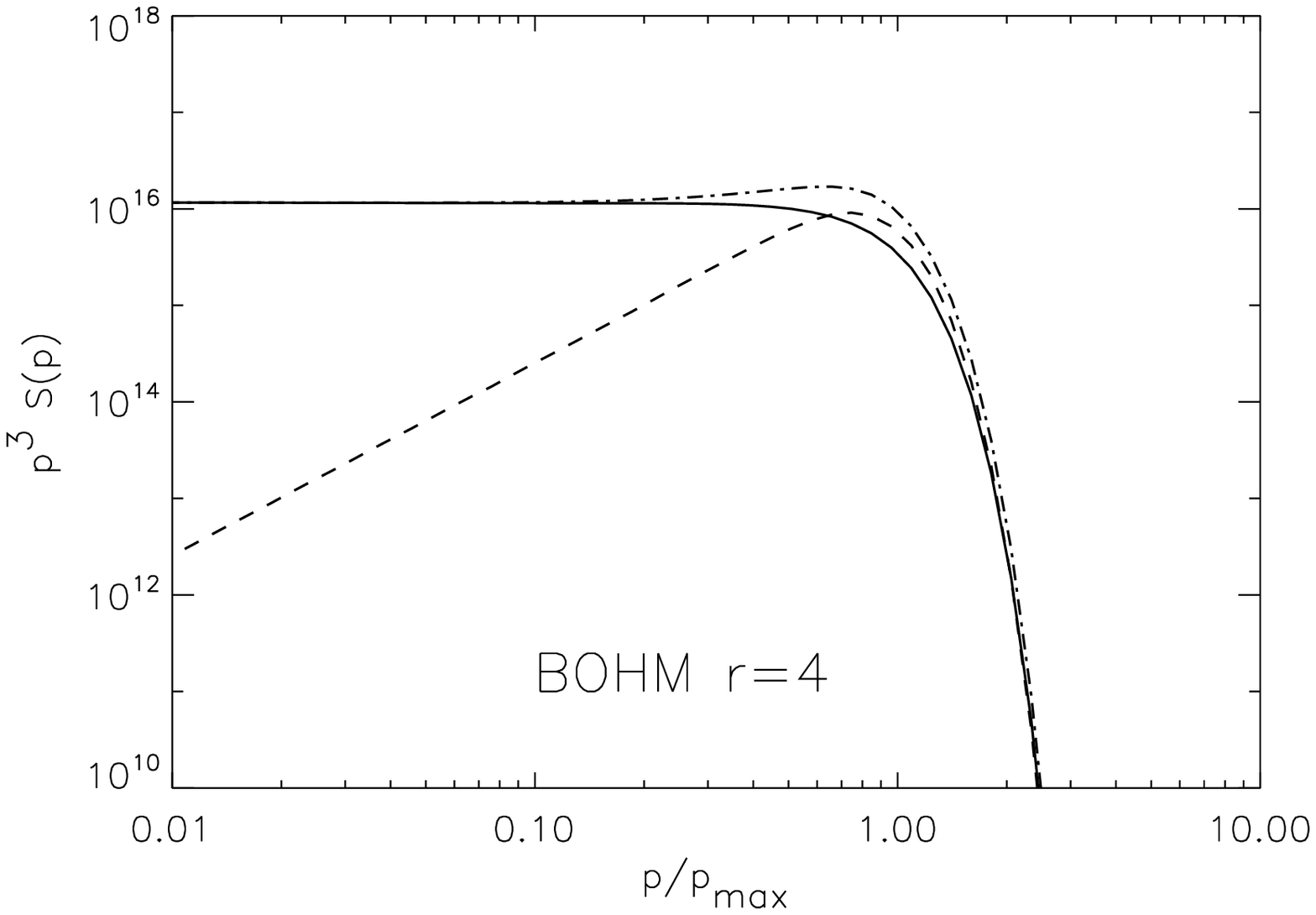}
\caption{Spectrum of accelerated particles for $r=4$ and Bohm difffusion, spatially integrated in the upstream fluid (dashed lines) and in the downstream fluid (solid lines). The left panel refers to the case in which no compression in the magnetic field is assumed, while the right panel shows the results for the case in which the magnetic field downstream is compressed by a factor $\sqrt{11}$ with respect to upstream. In both cases the momenta have been normalized to the same value of $p_{max}$ as calculated in the case without compression, in order to emphasize the fact that in the latter case the spectrum cuts off at a slightly lower momentum.}
\label{fig:spaceint}
\end{center}
\end{figure}

These expressions apply to any choice of the momentum dependence of the diffusion coefficient. The qualitative features of the integrated spectra upstream and downstream for $p<p_{max}$ can easily be understood. Let us for instance consider the case of Bohm diffusion: the upstream size of the region where particles with momentum $p$ diffuse is $\sim D(p)/u_{1}\propto p$. Therefore $S_{1}(p)\propto N_{0}(p)p\propto p^{-1}$ for $r=4$. The spatial size of the region where particles reside in the downstream region is $\sim u_{2}\tau_{loss,2}\propto 1/p$, therefore $S_{2}(p)\propto N_{0}(p)u_{2}\tau_{loss,2}\propto p^{-3}$ for $r=4$. Again, this trend is easily seen in the results of our exact calculations, shown in the left panel of Fig. \ref{fig:spaceint}. The solid line shows the spectrum as spatially integrated in the downstream region, while the upstream integrated spectrum is shown as a dashed line. The sum of the two is the dash-dotted line. Since $S_{1}(p)$ is much flatter, at most energies the largest contribution to the spectrum of radiating electrons in an astrophysical source comes from the downstream region. A small contribution appears only at $p\sim p_{max}$. This effect is even more visible if one assumes that the magnetic field downstream is obtained from the compression of the magnetic field in front of the shock surface, by an amount $\kappa^{-1}$. In the right panel of Fig. \ref{fig:spaceint} we show the case in which $\kappa^{-1}=\sqrt{11}$, which is the case of compression of an isotropic turbulent field upstream for $r=4$. One can see that in this case the downstream component produces a more pronounced bump in the total spectrum, as due to the fact that particles immediately downstream at $p\sim p_{max}$ lose energy more rapidly than upstream. While being hardly visible in the spectrum of synchrotron emission because this is dominated by the downstream region, this feature might be visible in the spectrum of inverse Compton radiation in those cases in which gamma ray emission can clearly be attributed to this process rather than to a hadronic origin. 

The simple trends illustrated above can be easily extended to any dependence of the diffusion coefficient on momentum. For instance for a constant diffusion coefficient, one can expect that $S_{1}(p)\propto p^{-2}$ and $S_{2}(p)\propto p^{-3}$. The slope of $S_{2}(p)$ is independent of the diffusion coefficient as it is determined by energy losses at all energies. Of course this simple prediction is the consequence of the assumption of stationarity of the problem. In real astrophysical sources, stationarity will be reached only at sufficiently high energies, while at lower energies the size of the downstream region is determined by the age of the source and is therefore independent of momentum. This is the reason for the appearance of spectral breaks in the spectra of radiation (see for instance \cite{karda}).

Another note of caution should be issued about the assumption of planar shock adopted throughout the calculations: in astrophysical sources the effects of sphericity might become important, especially on the volume integrated quantities. On the other hand one may argue that since particles only move a distance $\sim u_{2}\tau_{loss,2}(E)\propto 1/E$ downstream before losing an appreciable fraction of their energy, the effects induced by spherical symmetry are expected to become more important at low energies. It is likely however that the assumption of stationarity breaks down before the spherical geometry leads to important effects, but this should be checked case by case.

\section{Conclusions}
\label{sec:concl}

We proposed a semi-analytical calculation of the process of diffusive acceleration of electrons at a non-relativistic shock in the presence of synchrotron (or inverse Compton scattering) losses. The calculation returns the spectrum of electrons at the shock and at any location upstream and downstream for an arbitrary choice of the momentum dependence of the diffusion coefficient. The results of the calculations for the spectrum at the shock have been illustrated for three cases: 1) diffusion constant in momentum ($D(p)=D_{0}$), 2) Bohm diffusion ($D(p)\propto p$), and 3) Kolmogorov diffusion ($D(p)\propto p^{1/3}$). For the space dependence of the spectrum and the integrated electron spectra we restricted for simplicity to cases 1) and 2). 

While confirming the results of previous calculations for the simple case of a constant diffusion coefficient, our results can be easily obtained for any choice of the diffusion coefficient. Most results illustrated here are referred to the case of Bohm diffusion and to different compression factors at the shock. While the slope at low momenta is the standard test particle slope, depending only upon the compression factor, the shape of the cutoff, which is determined by the onset of synchrotron losses, depends on the adopted diffusion coefficient. It is exponential ($\propto \exp\left[ -p/p_{0}\right]$) for $D(p)=D_{0}$ and approaches a $\propto \exp\left[ -(p/p_{0})^{2}\right]$ for Bohm diffusion, where $p_{0}$ is related to $p_{max}$ through Eq. \ref{eq:p0pmax}.

The importance of these calculations for the description of the phenomenology of supernova remnants, and possibly other classes of sources, is evident: for supernova remnants the synchrotron X-ray emission is now resolved both spectrally and spatially and its careful description could allow to access precious information on the acceleration process. For instance, if the magnetic field is indeed amplified by accelerated particles (mainly protons) at the shock, the maximum energy of the electron component is determined by synchrotron losses, while if no amplification occurs, the maximum energy could be determined by the finite age (or spatial size) of the accelerator, as it is for protons. The two cases result in different shapes of the cutoff in the electron spectrum (exponential in the first case and super-exponential in the second case) and possibly in different fits to the synchrotron X-ray flux and morphology. 

One note of caution should be issued about the assumption of stationarity that is underling all calculations of this type. Strictly speaking, stationarity can be reached only if the time for energy losses is shorter than the age of the shock at all energies and the maximum energy is roughly obtained by equating the acceleration time with the loss time. In a real astrophysical source it is usually the case that at sufficiently low momenta the loss time is shorter than the age of the source, so that no stationarity can be actually reached at those momenta. In other words a spectral break can be expected in the volume integrated spectra of accelerated particles at some momentum $p_{b}$ and the stationary solution found here should be applied only for $p>p_{b}$ (e.g. \cite[]{karda}). This reflects in the spectra of the emitted radiation.

The formalism presented in this paper will be extended to the case of cosmic ray modified shocks in an upcoming paper: this application is crucial in that the presence of a precursor may appreciably flatten the electron spectra at high energy and lead to the production of more pronounced spectral  bumps.

\section*{Acknowledgments} 
The author is grateful to L. O'C. Drury for reading a preliminary version of the manuscript and E. Amato for a useful conversation on the spatial distribution of accelerated particles. This work was partially supported by MIUR (under grant PRIN-2006) and by ASI through contract ASI-INAF I/088/06/0. This research was also supported in part by the National Science Foundation under Grant No. PHY05-51164, in the context of the Program {\it Particle Acceleration in Astrophysical Plasmas}, July 26-October 3, 2009 help at the KITP in Santa Barbara. 

\appendix
\section{Green Function of the adjoint equation}

Here we briefly illustrate the determination of the Green function of the adjoint equation, defined by:
\begin{equation}
-u \frac{\partial \cG}{\partial x}-D(p)\frac{\partial^{2 \cG}}{\partial x^{2}}+Ap^{2}\frac{\partial \cG}{\partial p}=\delta(x-x')\delta(p-p'),
\end{equation}
where $\cG=\cG(x,p;x',p')$ and $A$ and $u$ are the appropriate quantities for the upstream and downstream regions. We introduce the following variables:
\begin{equation}
\xi = x-x'+\frac{u}{A}\left( \frac{1}{p'} - \frac{1}{p}\right)
\end{equation}
and
\begin{equation}
\tau = \frac{1}{A}\int_{p'}^{p}dy \frac{D(y)}{y^{2}}.
\end{equation}
In terms of these new variables, the equation above becomes:
\begin{equation}
-\frac{\partial^{2} \cG}{\partial \xi^{2}}+\frac{\partial \cG}{\partial \tau} =
\frac{\delta(\xi)\delta(\tau)}{Ap'^{2}}.
\end{equation}
Fourier transforming this equation with respect to $\xi$ one easily finds the solution in the form:
\begin{equation}
\cG(\xi)=\frac{1}{2\pi A p'^{2}} \left( \frac{\pi}{\tau}\right)^{1/2}
\exp\left[ -\frac{\xi^{2}}{4\tau}\right].
\end{equation}

\end{document}